\documentclass[aps,prl,preprintnumbers,nofootinbib,floatfix,onecolumn,12pt]{revtex4}

\usepackage{bm}
\usepackage{latexsym}
\usepackage{dcolumn}
\usepackage{amsfonts,amsmath,amssymb}
\usepackage{graphicx,epsfig}

\def	\be	{\begin{equation}}
\def	\ee	{\end{equation}}
\def	\d	{\delta}

\def	\lf	{\left (}
\def	\rt	{\right )}
\def	\del	{\nabla}

\begin{document}
%\title{Wald Entropy and the Einstein Equation of State}
\title{Beyond the Einstein Equation of State: Wald Entropy and Thermodynamical Gravity}

\author{Maulik Parikh and Sudipta Sarkar}

\affiliation{Inter-University Centre for Astronomy and Astrophysics (IUCAA), Post Bag 4, Pune 7, India}

\begin{abstract}
We show that the classical equations of gravity follow from a thermodynamic relation, $\delta Q = T \delta S$, where $S$ is taken to be the Wald entropy, applied to a local Rindler horizon at any point in spacetime. Our approach works for all diffeomorphism-invariant theories of gravity. This suggests that classical gravity may be thermodynamic in origin.
\end{abstract}

\maketitle

\section{Introduction}

Black holes have often provided insights into the nature of quantum gravity and the structure of spacetime. For example, the holographic principle emerged through Gedankenexperiments that took a property of black holes --- the sub-extensive scaling of Bekenstein-Hawking entropy --- and applied it to arbitrary gravitational systems. 

Another potentially profound insight comes from a paper by Jacobson \cite{einsteineqnofstate}. Jacobson considered the puzzling fact that the laws of black hole mechanics, derived in classical general relativity, seem mysteriously to anticipate the laws of black hole thermodynamics, derived in semi-classical gravity. Rather than trying to explain how classical laws could ``know about" quantum-mechanical ones, Jacobson reversed the logic, regarding the thermodynamics to be a premise rather than a consequence. Quite remarkably, by assigning the thermodynamic properties of black hole horizons to local light-cones in spacetime (not necessarily near a black hole), the Einstein equation re-appears as an equation of state. This seems to suggest, as indeed many theorists believe, that gravity is not a fundamental theory but originates in some kind of thermodynamic approximation.

The question arises whether this alluring result is somehow an artifact of Einstein gravity, or whether the connection between thermodynamics and gravity goes deeper, persisting also in general, higher-curvature theories of gravity. But extending the original derivation to higher-curvature theories is nontrivial, in part because that derivation makes use of the Raychaudhuri equation, whose usefulness is obscured in higher-curvature theories: the Raychaudhuri equation relates the derivative of the expansion of the horizon to the Ricci tensor, but a simple relation between the Ricci tensor and the stress tensor holds only for Einstein gravity. Moreover, in generic theories of gravity, the entropy is not simply proportional to the area.
%Indeed, in $f(R)$ gravity, the equations of motion do not seem to have a clear interpretation in terms of equilibrium thermodynamics.

In this paper, we obtain the classical gravitational equations from thermodynamics without making use of the Raychaudhuri equation. Specifically, 
%by focusing on the Lovelock class of gravitational theories --- a special class which includes Einstein's general theory of relativity as well as the higher-curvature terms in the low-energy limit of certain string theories --- 
we show that the classical equations of gravity follow directly from the Clausius relation, $\delta {\cal S} = \delta Q/T$. Here for ${\cal S}$ we use Wald's definition of entropy, which is the entropy (in place of $A/4$) that satisfies the first law of thermodynamics in higher-curvature theories. Our result suggests that
%, for Lovelock theories at least, 
classical gravity might have a quite intriguing thermodynamic origin.

\section{General Theories of Gravity}

Consider a general diffeomorphism-invariant theory of gravity in any number of dimensions. For convenience, we will assume that the Lagrangian is a polynomial in the Riemann tensor but does not involve its derivatives. One may regard the Lagrangian formally as dependent on both the metric and the Riemann tensor even though of course the Riemann tensor,  depends on the metric~\cite{Wald,Iyer}. Specifically, let the action be
\be
I = {1 \over 16 \pi} \int d^D x \sqrt{-g} L(g_{ab}, R_{abcd}) \label{action} \; .
\ee
We have set Newton's constant to unity. 
Define
\be
P^{abcd} = \frac{\partial L}{\partial R_{abcd}} \; .
\ee
$P^{abcd}$ has the same algebraic symmetries as the Riemann tensor, including cyclicity. 
One then finds that the equation of motion that follows from (\ref{action}) (supplemented by appropriate generalizations of Gibbons-Hawking-like boundary terms) is
\be
P_a^{\; cde} R_{bcde} - 2 \del^c \del^d P_{acdb} - \frac{1}{2} L g_{ab} = 8 \pi T_{ab} \; . \label{eom}
\ee
For example, when the Lagrangian is $L = f(R)$, we find $P^{abcd} = \frac{1}{2} f'(R) \lf g^{ac} g^{bd} - g^{ad} g^{bc} \rt$. Thus, the equation of motion is
\be
f'(R) R_{ab} - \del_a \del_b f'(R) + \left ( \Box f'(R) - \frac{1}{2} f(R) \right ) g_{ab} = 8 \pi T_{ab} \; .
\ee
This reduces to Einstein's equation when $f(R)= R$. 

Another example is Lovelock gravity~\cite{Lovelock,Lanczos}, the most general extension of Einstein gravity for which the equations of motion do not contain derivatives of the Riemann tensor. The Lagrangian is
%That is, the left-hand side of the equation of motion (the generalization of Einstein's equation) consists of a second-rank tensor (which generalizes the Einstein tensor) with terms like $R^2 g_{ab}$ and $R_{ac} R^c_b$, but not terms like $g_{ab} \Box R$ \cite{DeRuelle}. Unlike general higher-derivative theories, Lovelock gravity is ghost-free when expanded around flat space, thereby skirting around potential problems with unitarity; indeed, the graviton propagator is unaffected by the higher-derivative terms. Lovelock theories have attracted much attention in part because they include the first few terms in the $\alpha'$ expansion of heterotic and type I string theory \cite{Zwiebach,GrossWitten,BoulwareDeser}.
%The Lovelock Lagrangian consists of a finite series of terms,
$L = \sum_{m=0}^{m_{\rm max}} c_m L_m$,
where $c_m$ are constants of dimension $({\rm length})^{2m -2}$, which are arbitrary as far as gravity is concerned,  and $m_{\rm max} = (D-2)/2$ for even $D$ dimensions and $m_{\rm max} =(D-1)/2$ for odd $D$. Each term $L_m$ is made up of contractions of products of the Riemann tensor:
\be
L_m = {1 \over 2^m} \d^{i_1 ... i_{2m}}_{j_1 ... j_{2m}} R^{j_1 j_2}_{\quad \; i_1 i_2} ... R^{j_{2m -1} j_{2m}}_{\qquad \; \; i_{2m -1} i_{2m}} \; . \label{mthLagrangian}
\ee
Here the $\delta$ symbol is the generalized Kronecker delta, defined as the sum over signed permutations of products of ordinary Kronecker deltas. 
%For example,
%\begin{eqnarray}
%\d^{i_1 i_2}_{j_1 j_2} & = & \d^{i_1}_{j_1} \d^{i_2}_{j_2} - \d^{i_1}_{j_2} \d^{i_2}_{j_1} \nonumber \\
%$\d^{i_1 i_2 i_3}_{j_1 j_2 j_3}  = 
%\d^{i_1}_{j_1} \d^{i_2}_{j_2} \d^{i_3}_{j_3}
%+ \d^{i_1}_{j_2} \d^{i_2}_{j_3} \d^{i_3}_{j_1}
%+  \d^{i_1}_{j_3} \d^{i_2}_{j_1} \d^{i_3}_{j_2}
%-   \d^{i_1}_{j_1} \d^{i_2}_{j_3} \d^{i_3}_{j_2}
 %-   \d^{i_1}_{j_3} \d^{i_2}_{j_2} \d^{i_3}_{j_1}
 % -   \d^{i_1}_{j_2} \d^{i_2}_{j_1} \d^{i_3}_{j_3}
%$.
%\end{eqnarray}
%The generalized Kronecker delta is antisymmetric under interchange of any two upper indices as well as under interchange of any two lower indices, and thus vanishes when any two upper or lower indices are the same. This implies that $2 m_{\rm max} \leq D$, the dimension of spacetime. Actually, for even $D$ and $m = D/2$, the action (including the boundary term) is a topological invariant, the Euler character. Thus classically we find that $m_{\rm max} = (D-2)/2$ for even $D$ dimensions and $m_{\rm max} =(D-1)/2$ for odd $D$. %Thus, in $D=1,2$ only $L_0$ exists, in $D=3,4$ only $L_0$ and $L_1$ exist, in $D=5,6$, we have up to $L_2$, in $D=7,8$, we have up to $L_3$, and in $D=9,10$, we can also have a term $L_4$.
%The zeroeth term in the Lovelock Lagrangian is defined to be $L_0 = 1$ so that $c_0$ is proportional to the cosmological constant. The next term is
%\be
%L_1
% = {1 \over 2} ( \d^{i_1}_{j_1} \d^{i_2}_{j_2} - \d^{i_2}_{j_1} \d^{i_1}_{j_2} ) R^{j_1 j_2}_{i_1 i_2} 
%= R \; ,
%\ee
%the Ricci scalar. Thus, 
The Einstein-Hilbert action with a cosmological constant is just a special case of the Lovelock action with $c_1 = 1$ and $c_0 = - 2 \Lambda$. When $D \leq 4$, there are no other possible terms; the next term appears for $D \geq 5$. It is
%The Kronecker delta expansion has 24 (=4!) terms which, after contracting with the Riemann tensors, reduces to just three terms:
$L_2 
%=  {1 \over 4} \d^{i_1 i_2 i_3 i_4}_{j_1 j_2 j_3 j_4} R^{j_1 j_2}_{i_1 i_2} R^{j_3 j_4}_{i_3 i_4}
 = R^2 - 4 R^{ab} R_{ab} + R^{abcd} R_{abcd}$, known as Gauss-Bonnet gravity, which appears in the low-energy effective action of certain string theories~\cite{Zwiebach,GrossWitten}; its coefficient in ten-dimensional heterotic string theory is $c_2 = +\alpha'/4$. The Gauss-Bonnet action is a topological invariant in four dimensions, just as the Einstein-Hilbert action is a topological invariant in two dimensions.
%\be
%I = {1 \over 16 \pi} \int d^Dx \sqrt{-g} \left [ R - 2 \Lambda + c_2 (R^2 - 4 R^{ab} R_{ab} + R^{abcd} R_{abcd}) \right ]
%\ee
%There are also higher-curvature counterparts of the Gibbons-Hawking boundary term. 
%Black hole~\cite{BoulwareDeser,JTWheeler,MyersSimon,Cai} and cosmological solutions~\cite{Deruellecosmo,branecosmo} of Gauss-Bonnet gravity have been studied in some detail.
It is convenient to write (\ref{mthLagrangian}) in the form~\cite{PaddyAseem}
\begin{equation}
L_m =  Q^{a b c d}_{(m)} R_{ a b c d} \; . \label{lovelock}
\end{equation}
Then $P^{abcd} = m Q^{abcd}_{(m)}$, which has the nice property that $\del_a P^{abcd} = 0$. The equation of motion for Lovelock theory is
\begin{equation}
\sum_{m =0}^{m_{\rm max}} c_m \left ( m~ Q^{ a c d e}_{(m)} R^{b}_{\; c d e} -\frac{1}{2} L_m g^{ab} \right ) = 8 \pi T^{ab} \; , \label{eomLL}
\end{equation}
which follows easily from (\ref{eom}).
% find the following tensor
%\be
%\sigma^a_{(k)\, b} = - {1 \over 2^{k+1}} \d^{a i_1 ... i_{2k}}_{b j_1 ... j_{2k}} R^{j_1 j_2}_{i_1 i_2} ... R^{j_{2k -1} j_{2k}}_{i_{2k -1} i_{2k}} \; .
%\ee
%For instance when $m =1$ we have
%\begin{equation}
%Q^{a b c d}_{(1)} = \frac{1}{2} \left( g^{a c } g^{ b d} -  g^{a d } g^{ b c}\right) \; ,
%\end{equation}
%and hence for $c_m = \delta_{m,1}$ we obtain Einstein's equation:
%\be
%\sigma^a_{(1)\, b}
% = -{1 \over 4} \d^{a i_1 i_2}_{b j_1 j_2} R^{j_1 j_2}_{i_1 i_2} 
%R^{ab} - {1 \over 2} R g^{ab} = 8 \pi T^{ab} \; .
%\ee
%which of course is just Einstein's equation.
% Note that this can also be written as $R^{ab} - {1 \over 2} g^{ab} L_1$ where $L_1$ is the first Lovelock Lagrangian, $L_1 = R$.

%The next term, known as the Lanczos tensor, is
%\be
%\sigma^a_{(2)\, b} 
%=  -{1 \over 8} \d^{a i_1 i_2 i_3 i_4}_{b j_1 j_2 j_3 j_4} R^{j_1 j_2}_{i_1 i_2} R^{j_3 j_4}_{i_3 i_4}
% = 2(R R^a_b - 2 R^a_c R^c_b - 2 R^{cd} R^a_{c b d} + R^{a c d e} R_{b c d e}) - {1 \over 2} \d^a_b L_2
%\ee
%The general Lovelock tensor (with both indices raised or lowered) is symmetric under interchange of $a$ and $b$. In particular, expanded around flat space, all the higher derivative terms vanish since the coefficients are all zero. The Lovelock tensors are also divergence-free. To show that one takes the divergence of the Lovelock tensor and then uses the Bianchi identity: $\del_\sigma R_{\mu \nu \tau \xi} + \del_\tau R_{\mu \nu \xi \sigma} + \del_\xi R_{\mu \nu \sigma \tau} = 0$.

In each of these theories, one can associate an entropy with Killing or black hole horizons. For example, in place of $A/4$, the entropy in $f(R)$ gravity is
\be
{\cal S}_{f} = f'(R) \frac{A}{4} \; , \label{entropyf}
\ee
while for Gauss-Bonnet gravity, black holes have an entropy of
\be
{\cal S}_{\rm G-B} = \frac{1}{4} \int d^{D-2} x \sqrt{\sigma} \left ( 1 + 2 c_2 \, {}^{(D-2)}R \right ) \; ,\label{entropygb}
\ee
where ${}^{(D-2)}R$ is the scalar curvature of (the cross-section of) the horizon. We will show below that, as in Jacobson's derivation of Einstein's equation from $S = A/4$~\cite{einsteineqnofstate}, varying these entropies and imposing the Clausius relation, $\delta Q = T \delta S$, leads directly to the equations of classical gravity.

\section{Wald Entropy}

%\def\eq#1{{Eq.~(\ref{#1})}}
%\def\mes#1#2{\int_{#2} d^{#1}x\, \sqrt{-g}\, }

%We review the concept of entropy associated with a spacetime horizon in a general theory of gravity.
Wald \cite{Wald,Iyer} and other authors \cite{MyersKangJac, JacobMyers} have developed a powerful and elegant Lagrangian-based method for determining the entropy of a black hole with a Killing horizon. Wald's method works for any diffeomorphism-invariant theory in any number of dimensions and does not require Euclideanization. 
%The formalism is due to Wald and collaborators \cite{Wald} and further developed by authors in \cite{Iyer, MyersKangJac, JacobMyers}. 
Here we adopt a simplified version of the formalism \cite{Cardoso}. Consider a generally covariant Lagrangian, $L$, that
depends on the Riemann tensor but does not contain derivatives
of the Riemann tensor. 
%The entropy is interpreted as the Noether charge associated with diffeomorphism invariance. We shall brieﬂy summarize
%this approach and use this definition.
%To define the Noether charge associated with the diffeomorphism invariance, let us consider the 
Under the diffeomorphism $x^a \rightarrow x^a + \xi^a$ the metric changes via $\delta g_{ab} = -\del_a \xi_b - \del_b \xi_a$. By diffeomorphism-invariance, the change in the action, when evaluated on-shell, is given only by a surface term. 
%[The subscript $\xi$ on $\delta_\xi....$ is a reminder that we are considering the changes due to a particular kind of variation, viz. when the metric changes by $\delta g_{ab} = - (\del_a \xi_b + \del_b \xi_a)$.] 
%Here $L$ is any diffeomorphism invariant Lagrangian and $V^a$ is a $D$ dimensional object (\textit{not a vector}!) which is related with the surface term. 
This leads to a conservation law, $\del_a J^a =0$, for which we can write $J^a =  \del_b J^{ab}$, where $J^{ab}$ defines (not uniquely) the antisymmetric Noether potential associated with the diffeomorphism $\xi^a$ \cite{Wald}.

For a Lagrangian of the type $L=L(g_{ab},R_{abcd})$ direct computation shows that $J^{ab}$ is given by (see \cite{Cardoso})
\begin{equation}
J^{ab} = - 2 P^{abcd} \del_c \xi_d + 4 \xi_d \left(\del_c P^{abcd}\right) \; ,
\label{noedef}
\end{equation} 
with $P^{abcd} = \partial L/\partial R_{abcd}$. The Noether charge associated with a rigid diffeomorphism $\xi^a$ is defined by integrating the Noether potential over a closed spacelike surface $S$:
\begin{equation}
Q = \int_{S} J^{a b} dS_{a b} \; .
\end{equation}
When $\xi^a$ is a timelike Killing vector (the one whose norm vanishes at the Killing horizon), it turns out \cite{Wald,Iyer} that the corresponding Noether charge is precisely the entropy, ${\cal S}$, associated with the horizon, apart from a few factors:
\begin{equation}
{\cal S} = \frac{1}{8 \kappa}  \int_S d S_{a b} J^{a b} \; . \label{generalform}
\end{equation}
Here $\kappa$ is the surface gravity of the black hole horizon. The integral for this ``Wald entropy" can be evaluated over any spacelike cross-section of the Killing horizon \cite{MyersKangJac}.
%Note that for most general Lagrangians, this integral has to be taken over the bifurcation surface so that all the terms in the Noether charge proportional to $\xi^a$ vanishes and the final form becomes very simple. The standard procedure is then to express $dS_{a b}$ in terms of the \textit{bi-normals}, $\epsilon_{a b}$ and then to use the relationship $ \del_a \xi_b =\kappa \epsilon_{a b} $ (which is valid only on the bifurcation surface) to write \eq{En:generalform} in the familiar form \cite{Wald,Iyer}. For our purpose, we would like to work with the general result in \eq{En:generalform}. 
In fact we can formally define the quantity ${\cal S}$ on any closed spacelike surface, $S$, of codimension two (such as a section of a stretched horizon), and only at the end take the limit in which that $S$ approaches a section of the Killing horizon. It can be shown, for example, that both (\ref{entropyf}) and (\ref{entropygb}) are just special cases of Wald entropy.
%But, ${\cal S}$ has the interpretation as the entropy only when $dS_{a b}$ is the surface element on the horizon.
%Wald's definition of entropy satisfies the first law of thermodynamics in general theories of gravity and, indeed, nicely explains how $\delta M$, a variation at infinity, can be proportional to $\delta S$, a variation at the horizon: both are now defined with respect to $\xi^a$.

\section{Gravitation From Thermodynamics}

Now let us show how the classical equations of gravity, (\ref{eom}), arise thermodynamically. (That the equations look thermodynamical has been shown for spherically-symmetric Lovelock gravity~\cite{AseemSudipta}.) The set-up is as follows \cite{einsteineqnofstate}.
%Take a spacetime point $p$. Let $B_1$ be any spacelike neighborhood of $p$ of codimension two. We let $B_1$ be contained in a slightly larger, ``stretched" spacelike surface $S_1$. $B_1$ and $S_1$ are both taken to be small enough that the spacetime in their vicinity can be approximated as flat Minkowski space. In particular, there is a future-directed timelike Killing vector field $\xi^a$ that generates boosts orthogonal to $S_1$. The orbits of this Killing vector field emanating from $S_1$ trace out the stretched horizon $\Sigma$, a timelike surface of codimension one. See Fig. 1. At each point on $\Sigma$ there is a spacelike vector $n^a$ that is normal to $\Sigma$. Similarly, from $B_1$ we send out future-directed null rays, $k^a$, orthogonal to $B_1$ and such that their expansion is non-negative. These generates the true local Rindler horizon, $H$, near $p$. 
Take any spacetime point $p$ and pick any future-directed null vector $k^a$ emanating from $p$. In the vicinity of $p$, the plane orthogonal to $k^a$ defines a local acceleration, or Rindler, horizon, $H$. Let $B_1$ be any spacelike neighborhood of $p$ of codimension two that locally lives on the Rindler plane, and let $B_2$ be some further section of the Rindler plane along $k^a$. Next, let $\xi^a$ be a future-directed approximate timelike Killing vector that generates boosts and asymptotically approaches $k^a$. The orbits of $\xi^a$ and the plane orthogonal to the acceleration vector of $\xi^a$ define a stretched horizon, $\Sigma$. As in the membrane paradigm \cite{membraneparadigm,membrane}, points on $H$ and points on $\Sigma$ can be put in one-to-one correspondence by, say, ingoing null rays that pierce both surfaces. Let $S_i$ be the images of $B_i$ on $\Sigma$ via this correspondence. See Fig. 1.

Let $\xi^a \xi_a = -\alpha^2$, where the norm $\alpha$ (which turns out to be a lapse) is taken to be constant over $\Sigma$. This norm vanishes at $H$, a Killing horizon. Let $u^a$ be the proper velocity of a fiducial observer moving along the orbit of $\xi^a$ i.e. $u^a = \left ( \frac{d}{d \tau} \right )^{\! a} = \frac{1}{\alpha} \xi^a$, where $\tau$ is the proper time. Let $n^a$ be the spacelike unit normal to $\Sigma$, pointing in the direction of increasing $\alpha$. Both $u^a$ and $n^a$ map to $k^a$ in the limit that $\alpha \to 0$, for which $\Sigma \to H$.

\begin{figure}[hbtp]
 \centering
  \epsfysize=5 cm
  \epsfbox{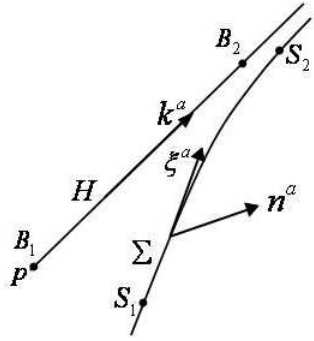}
 \centering
 \caption{The local Rindler horizon, $H$, of an arbitrary spacetime point $p$ is defined by a null vector, $k^a$. A stretched horizon, $\Sigma$, is defined by a timelike approximate Killing vector $\xi^a$ and has a normal vector field $n^a$. $B_i$ and $S_i$ are spacelike patches of codimension two that inhabit the planes of the Rindler and stretched horizons in the directions orthogonal to the figure, with $p$ contained in $B_1$.}
\end{figure}

%So far, we have considered a general Lagrangian which depends on curvature tensor. Now, we like to concentrate on a particular subclass which is the case of Lovelock Lagrangians. 
%As we mentioned, these Lagrangians can be written as in the form,
%\begin{equation}
%L = \frac{1}{16 \pi } Q^{a b c d} R_{ a b c d}. \label{lovelock}
%\end{equation}
%The tensor $Q^{a b c d}$ is divergenceless in all indices and it also has all the symmetries of curvature tensor. For $m$-th order lovelock case, $L$ is a homogeneous function of degree $m$.
After these preliminaries, we are ready to deduce the classical equations of gravity from thermodynamics. The key idea \cite{einsteineqnofstate} is to assign black hole thermodynamic properties to local Rindler horizons \cite{paddyinside}. The stretched horizon is assigned a local temperature, $T_{\rm loc} = \kappa/2 \pi \alpha$, as well as the Wald entropy appropriate to the given theory of gravity. 

By (\ref{noedef}) and (\ref{generalform}), the Wald entropy associated with a stretched horizon at time $\tau$ is
\be
{\cal S} 
%= \frac{ \pi}{ \kappa} \int_{S} dS_{ab} ~J^{ab} 
= - \frac{1}{4 \kappa} \int_{S(\tau)} dS_{ab}~ \left ( P^{a b c d} \del_{c}\xi_{d} - 2 \xi_d \del_c P^{abcd} \right ) \; .
\ee
%, then on the $(D-2)$ dimension cross section of the horizon, it is possible to write $d\Sigma_{a b } = \epsilon_{a b} \sqrt{\sigma} d^{D-2}x $ and using the relation $ \epsilon_{a b} = \kappa \del_{a}\xi_{b}$ valid on the bifurcation surface, one obtains the usual form of entropy. The normalization is fixed so that it gives correct value for Einstein case namely the one-quarter of area.
Next, we vary the entropy along the timelike congruence. The entropy change is
% TWO COLUMN ARRAY BELOW
%\begin{eqnarray}
%\delta {\cal S} & = & {\cal S}(\tau_2)- {\cal S}(\tau_1) \nonumber \\ 
%&=&  - \frac{1}{4 \kappa}  \left [ \int_{S(\tau_2)} dS_{ab}~ \left ( P^{a b c d} \del_{c}\xi_{d} - 2 \xi_d \del_c P^{abcd} \right ) \right . \nonumber \\
%& & \left . \qquad - \int_{S(\tau_1)} dS_{ab}~ \left ( P^{a b c d} \del_{c}\xi_{d} - 2 \xi_d \del_c P^{abcd} \right ) \right ] \nonumber \\
%- \frac{m}{4 \kappa}\left[ \int_{S(\tau_2)} \! \! \! \! \! \! \! \! dS_{a b } \, Q^{a b c d}_{(m)}\del_{c}\xi_{d}(\tau_2) - \! \!\int_{S(\tau_1)} \! \! \! \! \! \! \! \! dS_{a b} \, Q^{a b c d}_{(m)} \del_{c}\xi_{d}(\tau_1)\right] \nonumber \\
%&=&  + \frac{1}{4 \kappa} \int_{\Sigma} d\Sigma_{a} \del_{b}
%\left(P^{a b c d} \del_{c}\xi_{d} - 2 \xi_d \del_c P^{abcd} \right )
%Q^{a b c d}_{(m)}\del_{c}\xi_{d}\right) \; .
%&=&  - \frac{m}{4 \kappa} \int_{\Sigma} d\Sigma_{b}Q^{a b c d}_{(m)} R_{ d c a m} \xi^{m} \; .
%\end{eqnarray}
% ONE COLUMN ARRAY BELOW
\begin{eqnarray}
\delta {\cal S} & = & {\cal S}(\tau_2)- {\cal S}(\tau_1) \nonumber \\ 
&=&  - \frac{1}{4 \kappa}  \left [ \int_{S(\tau_2)} dS_{ab}~ \left ( P^{a b c d} \del_{c}\xi_{d} - 2 \xi_d \del_c P^{abcd} \right ) - \int_{S(\tau_1)} dS_{ab}~ \left ( P^{a b c d} \del_{c}\xi_{d} - 2 \xi_d \del_c P^{abcd} \right ) \right ] \nonumber \\
%- \frac{m}{4 \kappa}\left[ \int_{S(\tau_2)} \! \! \! \! \! \! \! \! dS_{a b } \, Q^{a b c d}_{(m)}\del_{c}\xi_{d}(\tau_2) - \! \!\int_{S(\tau_1)} \! \! \! \! \! \! \! \! dS_{a b} \, Q^{a b c d}_{(m)} \del_{c}\xi_{d}(\tau_1)\right] \nonumber \\
&=&  + \frac{1}{4 \kappa} \int_{\Sigma} d\Sigma_{a} \del_{b}
\left(P^{a b c d} \del_{c}\xi_{d} - 2 \xi_d \del_c P^{abcd} \right )
%Q^{a b c d}_{(m)}\del_{c}\xi_{d}\right) \; .
%&=&  - \frac{m}{4 \kappa} \int_{\Sigma} d\Sigma_{b}Q^{a b c d}_{(m)} R_{ d c a m} \xi^{m} \; .
\end{eqnarray}
In the last step, we have used Stokes' theorem for an antisymmetric tensor field $A^{ a b}$:
\begin{eqnarray}
\int_{\Sigma} d\Sigma_{a} \del_b A^{ a b} = -\oint_{S} dS_{a b } A^{a b} \; ,
\end{eqnarray}
where our $\Sigma$ has the boundary $S = S(\tau_1) \cup S(\tau_2) $, and the minus sign comes about because $\Sigma$ is timelike. (To be explicit, our conventions here are $d\Sigma_a = n_a \, dA\, d \tau$ and $dS_{ab} = \frac{1}{2} (n_a u_b - u_a n_b) dA$, where the normal $n^a$ to the stretched horizon points outwards, away from the true horizon.) Recall that $P^{abcd}$ has the same algebraic symmetries as the Riemann tensor, including cyclicity. Using those symmetries, we find that
% TWO COLUMN ARRAY BELOW
%\begin{eqnarray}
%\delta {\cal S} & = & \frac{1}{4 \kappa} \int_\Sigma \left [- \del_b \left (P^{adbc} + P^{acbd} \right ) \del_c \xi_d \right . \nonumber \\
%&  & + P^{abcd} \del_b \del_c \xi_d - \left . 2 \xi_d \del_b \del_c P^{abcd}  \right ] d \Sigma_a
%\end{eqnarray}
% ONE COLUMN ARRAY BELOW
\be
\delta {\cal S} = \frac{1}{4 \kappa} \int_\Sigma \left [- \del_b \left (P^{adbc} + P^{acbd} \right ) \del_c \xi_d + P^{abcd} \del_b \del_c \xi_d - 2 \xi_d \del_b \del_c P^{abcd}  \right ] d \Sigma_a
\ee
So far we have not used any properties of $\xi^a$. Now we will assume that $\xi^a$ is an approximate timelike Killing vector. An exact Killing vector satisfies Killing's equation $\del_b \xi_c + \del_c \xi_b = 0$ from which it follows that $\del_a \del_b \xi_c = R^d_{\; abc} \xi_d$. An approximate Killing vector indeed satisfies Killing's equation locally. We will also assume the applicability of the second equation within our local Rindler patch, a point we will discuss in the next section. The terms in parentheses drop out by symmetry. Using our assumption, we find
\be
T_{\rm loc} \delta {\cal S} = \frac{1}{8 \pi \alpha} \int_\Sigma\! \! \!  \lf  P^{abcd} R_{dcbe} \xi^e \! - 2 \xi_d \del_b \del_c P^{abcd} \rt n_a d \tau d A
\ee
%Then, using the divergencelessness of $Q^{abcd}_{(m)}$, the Killing identity $\del_b \del_c \xi_d = R^e_{\; bcd} \xi_e$, and the local Rindler temperature $T_{\rm loc} = \frac{\kappa}{2 \pi \alpha}$, we find
%\begin{equation}
%\delta {\cal S} =  - \frac{m}{4 \kappa} \int_{\Sigma} d\Sigma_{b}Q^{a b c d}_{(m)} R_{ d c a m} \xi^{m} \; .
%\end{equation}
%the $T \delta {\cal S}$ side can be written as
%Multiplying by the local temperature, $T_{\rm loc}$, we find
%\begin{equation}
%T_{\rm loc} \delta {\cal S} = \frac{m}{8 \pi \alpha}\int_{\Sigma} d\Sigma_{a} Q^{a b c d} _{(m)} R_{ebcd} \xi^{e} \; .
%\end{equation}
On the other hand, the locally-measured energy or heat flux into the stretched horizon is
\begin{equation}
\delta Q = + \int_\Sigma d \Sigma_a T^a_e u^e = \frac{1}{\alpha} \int_\Sigma dA \, d \tau \, n_a T^a_e \xi^e \; .
\end{equation}
%Already at this level if we were to apply the Clausius relation, $T_{\rm loc} \delta {\cal S} = \delta Q$, we would arrive at the correct equations of motion, (\ref{eom}). However, 
Now ${\cal S}$ is not yet the entropy of the true horizon, $H$, since we still have to take the limit in which the stretched horizon becomes null. Then both $n^a$ and $u^a$ become proportional to the null vector $k^a$ (with the same proportionality constant). Equating $\delta Q$ and $T_{\rm loc} \delta {\cal S}$ and taking the null limit, we obtain
%\begin{equation}
%T \delta {\cal S} = 
%\frac{m \kappa }{8 \pi} \int_H d\lambda \, \lambda \, dA \, Q^{a b c d}_{(m)} R_{ebcd} k_a k^e
%= \kappa \int_H d\lambda \, \lambda \, dA T^{a}_{e} k_a k^e \; . \label{clausius}
%\end{equation}
%where we have used the relations $ \xi^a = \lambda \kappa k^a $ and $d\Sigma_{b} = k_b d\lambda dA$.
%We now take the limit $\alpha \to 0$.
%\begin{equation}
%\delta Q = \kappa \int d\lambda~ \lambda ~ dA~ T^{b}_{m} k_b k^m \; .
%\end{equation}
%Hence, the Clausius relation $T \delta S = \delta Q$ immediately gives
\be
\lf P_a^{\; cde} R_{bcde} - 2 \del^c \del^d P_{acdb} \rt k^a k^b = 8 \pi T_{ab} k^a k^b \; .
\ee
%If we now require that $\delta {\cal S} = \delta Q/T$ hold for all local Rindler
%horizons through all points $p$, we infer that the two integrands in (\ref{clausius}) must agree 
Since this holds for all null vectors $k^a$ at $p$, we infer that
\begin{equation}
P_a^{\; cde} R_{bcde} - 2 \del^c \del^d P_{acdb} + \varphi \, g_{ab} = 8 \pi T_{ab} \; ,
\end{equation}
for some scalar function $\varphi$. By demanding conservation of the stress tensor and using the Bianchi identities, we find that $\varphi = - \frac{1}{2} L + \Lambda$, where $\Lambda$ is an integration constant. Thus we see that imposing $T_{\rm loc} \delta {\cal S} = \delta Q$ at any point in spacetime necessarily implies that
\begin{equation}
P_a^{\; cde} R_{bcde} - 2 \del^c \del^d P_{acdb} -\frac{1}{2} L g_{ab} + \Lambda g_{ab} = 8 \pi T_{ab} \; .
\end{equation}
With the cosmological constant appearing as an integration constant, this is precisely the classical equation of motion, (\ref{eom}), for our theory of gravity.
%In particular, when $m = 1$ we have 
%\begin{equation}
%Q^{a b c d} = \frac{1}{2} \left( g^{a c } g^{ b d} -  g^{a d } g^{ b c}\right) \; .
%\end{equation}
%and (\ref{eom}) becomes Einstein's equations. For $m = 2$, we recover the gravitational equations for pure Gauss-Bonnet gravity.

\section{Discussion}

We have shown that the equations of classical gravity follow from thermodynamics. Our derivation did not require the Raychaudhuri equation. Moreover, since we started with the Wald entropy, we could go beyond the Einstein equation to the equations of motion of general theories of gravity.
%the higher-curvature Lovelock equations. 
Since these were obtained from the Clausius relation, they can be regarded as equations of state --- relations between thermodynamic state variables.

%The question arises whether this result extends to all theories of gravity. In general, the divergence of $\partial L / \partial R_{abcd}$ does not vanish, as it does for Lovelock theories, and the second term in (\ref{noedef}) also contributes to the Noether potential and the entropy. It is unclear, however, whether the derivation goes through even when this additional term is incorporated.
Satisfying as this is, there remain some loose ends. One observation~\cite{observation} is that the derivation of the Wald entropy itself relies on the equations of motion being obeyed. Although our approach never explicitly invokes the equations of motion, it is still unclear whether any derivation, including Jacobson's original calculation, that begins with an on-shell expression which agrees with Wald entropy (such as $A/4$) is implicitly assuming the answer, or whether that is simply self-consistency. Another technical concern is our use of the equation $\del_a \del_b \xi_c = R_{abcd} \xi^d$. This equation is obviously true when $\xi^a$ is an exact Killing vector but to what extent can one trust it when $\xi^a$ is an approximate Killing vector? For a general spacetime, Riemann normal coordinates can be applied to any local patch. In such coordinates an approximate boost Killing vector looks like $x \partial_t + t \partial_x$. However, such a vector does not obey  $\del_a \del_b \xi_c = R_{abcd} \xi^d$ in general, so our assumption cannot be satisfied through coordinate choices alone. In this light, it is interesting that for $f(R)$ theories, previous work has found that the Clausius relation does not give the equations of motion but also has additional terms~\cite{jacobsonf(R)}; these have been interpreted as non-equilibrium effects. Perhaps the failure of  $\del_a \del_b \xi_c = R_{abcd} \xi^d$ to hold may be traced to such effects. In that case, our derivation may indicate how to determine potential non-equilibrium terms for a general theory of gravity. (On the other hand, we also make use of this equation for Einstein gravity where there are no such terms, so perhaps it is an innocuous assumption.) It would be interesting to understand this better, as well as to connect our method to previous approaches~\cite{jacobsonf(R)}.
%But if all gravitational theories cannot be derived thermodynamically, what characterizes those theories that can?
%In this context, it would be interesting to check whether the equations for fields related to the graviton --- such as the dilaton and the gravitino --- also have a thermodynamic origin.

While this paper was being prepared, the preprint~\cite{brustein} appeared, claiming similar conclusions; unfortunately, among other things, their starting formula for entropy (equation 9 in~\cite{brustein}) is manifestly incorrect, leaving the result in doubt.

\vspace{0.3cm}

\noindent
{\bf Acknowledgments}

\noindent
We would like to thank T. Padmanabhan for discussions.
S. S. is supported by the Council of Scientific and Industrial Research, India.

\end{document}